\documentclass[12pt]{article} \usepackage{amssymb}
\begin {document}


\def\bbbf{{\rm I\!F}}

\def\bbbz{{\mathchoice {\hbox{$\sf\textstyle Z\kern-0.4em Z$}}
{\hbox{$\sf\textstyle Z\kern-0.4em Z$}}
{\hbox{$\sf\scriptstyle Z\kern-0.3em Z$}}
{\hbox{$\sf\scriptscriptstyle Z\kern-0.2em Z$}}}}

\newenvironment{proof}{\begin{trivlist}\item[]{\em Proof: }}%
{\samepage \hfill{\hbox{\rlap{$\sqcap$}$\sqcup$}}\end{trivlist}}

\newtheorem{definition}{Definition}

\newtheorem{theorem}{Theorem}
\newtheorem{lemma}{Lemma}
\newtheorem{corollary}{Corollary}
\newtheorem{remark}{Remark}
\newtheorem{example}{Example}

\title{Reducible Cyclic Codes Constructed as the Direct Sum of Two Semiprimitive Cyclic Codes}

\author{
Gerardo Vega\thanks{This work was partially supported by PAPIIT-UNAM IN107515.}\thanks{G. Vega is with the Direcci\'on General de C\'omputo y de Tecnolog\'{\i}as de Informaci\'on y Comunicaci\'on, Uni\-ver\-si\-dad Nacional Aut\'onoma de M\'exico, 04510 M\'exico D.F., MEXICO (e-mail: gerardov@unam.mx).}}
\markboth{IEEE Transactions on Information Theory}{G. Vega: The Weight Distribution of a Generalized Family of Reducible Cyclic Codes}
\maketitle

\begin{abstract} 
We present a family of reducible cyclic codes constructed as the direct sum of two different semiprimitive two-weight irreducible cyclic codes. This family generalizes the class of reducible cyclic codes that was reported in the main result of B. Wang, {\em et al.} \cite{once}. Moreover, despite of what was stated therein, we show that, at least for the codes studied here, it is still possible to compute the frequencies of their weight distributions through the cyclotomic numbers in a very easy way.
\end{abstract}


\section{Introduction}
It is said that a cyclic code is reducible if its parity-check polynomial is factorizable in two or more irreducible factors. Each one of these irreducible factors can be seen as the parity-check polynomial of an irreducible cyclic code. Therefore, a reducible cyclic code is, basically, the direct sum of these irreducible cyclic codes. Reducible cyclic codes whose parity-check polynomials are factorizable in exactly two different irreducible factors have been extensively studied (see, for example, \cite{cuatro}, \cite{doce}, \cite{siete}, \cite{seis}, \cite{dos}, \cite{once}, \cite{trece} and \cite{nueve}). A very interesting problem regarding this kind of reducible cyclic codes is to obtain their full weight distributions. In particular, B. Wang, {\em et al.} \cite{once} employed an elaborate procedure that uses some sort of elliptic curves in order to obtain the weight distribution of a class of reducible cyclic codes. We present here a family of reducible cyclic codes constructed as the direct sum of two different semiprimitive two-weight irreducible cyclic codes, that generalizes such class of reducible cyclic codes. Moreover, we show that, contrary to what was stated in \cite[p. 7254]{once}, it is still possible, at least for the codes in this family, to compute the frequencies of their weight distributions through the cyclotomic numbers in a very easy way.

\begin{center}
TABLE I \\
{\em Weight distribution of ${\cal C}_{(a_i)}$, $i=1,2$. \\}
\end{center}
\begin{center}
\begin{tabular}{|c|c|} \hline
{\bf Weight} & $\;$ {\bf Frequency} $\;$\\ \hline \hline
0 & 1 \\ \hline
$\; \frac{\lambda}{2} (q^{k-1}-q^{(k-2)/2}) \;$ & $\frac{(q^k-1)}{2}$ \\ \hline
$\; \frac{\lambda}{2} (q^{k-1}+q^{(k-2)/2}) \;$ & $\frac{(q^k-1)}{2}$ \\ \hline
\end{tabular}
\end{center}

In order to give a detailed explanation of what is the main result of this work, let $p$, $t$, $q$, $k$ and $\Delta$ be integers, such that $p$ is a prime, $q=p^t$ and $\Delta=(q^k-1)/(q-1)$. In addition, let $\gamma$ be a fixed primitive element of $\bbbf_{q^k}$ and, for any integer $a$, denote by $h_a(x) \in \bbbf_{q}[x]$ the minimal polynomial of $\gamma^{-a}$. With this notation, the following result gives a description for the weight distribution of a family of reducible cyclic codes:

\begin{center}
TABLE II \\
{\em Weight distribution of ${\cal C}_{(a_1,a_2)}$.}
\end{center}
\begin{center}
\begin{tabular}{|c|c|} \hline
{\bf Weight} & $\;$ {\bf Frequency} $\;$\\ \hline \hline
0 & 1 \\ \hline
$\; \frac{\lambda}{3} (q^{k-1}-q^{(k-2)/2}) \;$ & $\frac{3(q^k-1)}{2}$ \\ \hline
$\; \frac{\lambda}{3} (q^{k-1}+q^{(k-2)/2}) \;$ & $\frac{3(q^k-1)}{2}$ \\ \hline
$\; \frac{\lambda}{2} (q^{k-1}-q^{(k-2)/2}) \;$ & $\frac{(q^k-1)(q^k-5)}{8}$ \\ \hline
$\; \frac{\lambda}{6} (3q^{k-1}-q^{(k-2)/2}) \;$ & $\frac{3(q^k-1)^2}{8}$ \\ \hline
$\; \frac{\lambda}{6} (3q^{k-1}+q^{(k-2)/2}) \;$ & $\frac{3(q^k-1)^2}{8}$ \\ \hline 
$\; \frac{\lambda}{2} (q^{k-1}+q^{(k-2)/2}) \;$ & $\frac{(q^k-1)(q^k-5)}{8}$ \\ \hline
\end{tabular}
\end{center}

\begin{theorem}\label{teouno}
Suppose that $3 | (q^k-1)$ and $3 \nmid \Delta$. Let $a_1$, $a_2$, $n$ and $a$ be any integers such that $a_1-a_2=\pm \frac{q^k-1}{3}$, $n=\max\{\frac{q^k-1}{\gcd(q^k-1,a_i)} \}_{i=1}^2$ and let $a=a_1$, if $n=\frac{q^k-1}{\gcd(q^k-1,a_1)}$; otherwise let $a=a_2$. Also suppose that $\gcd(\Delta,a)=2$, and fix $\lambda$ such that $\gcd(q-1,\frac{a}{2})=\frac{q-1}{\lambda}$. If ${\cal C}_{(a_1)}$, ${\cal C}_{(a_2)}$ and ${\cal C}_{(a_1,a_2)}$ are, respectively, the cyclic codes with parity-check polynomials $h_{a_1}(x)$, $h_{a_2}(x)$ and $h_{a_1}(x)h_{a_2}(x)$, then 

\begin{enumerate}
\item[{\em (A)}] ${\cal C}_{(a_1)}$ and ${\cal C}_{(a_2)}$ are two different semiprimitive two-weight irreducible cyclic codes of length $n$ and dimension $k$. In addition, these codes have the same weight distribution which is given in Table I.
 
\item[{\em (B)}] ${\cal C}_{(a_1,a_2)}$ is an $[n,2k]$ cyclic code over $\bbbf_{q}$, with the weight distribution given in Table II.
\end{enumerate}
\end{theorem}

Recently, G. Vega \cite{nueve} gave a unified explanation for the weight distribution of several families of codes whose parity check polynomials are given by the products of the form $h_{a}(x)h_{a \pm \frac{q^k-1}{2}}(x)$, where $h_{a}(x) \neq h_{a \pm \frac{q^k-1}{2}}(x)$. From this perspective, therefore, it is important to keep in mind that the parity check polynomials of the kind of codes studied by B. Wang, {\em et al.} \cite{tres}, and those studied by Theorem \ref{teouno}, are now given by the products of the form $h_{a}(x)h_{a \pm \frac{q^k-1}{3}}(x)$.

This work is organized as follows: In Section 2 we establish some notation, recall some definitions and establish our main assumption. Section 3 is devoted to presenting some preliminary and general results. In Section 4 we use these results in order to present a formal proof of Theorem \ref{teouno}, and show that the class of reducible cyclic codes studied in \cite{once}, can be obtained as particular instances of Theorem \ref{teouno}. Finally, Section 6 will be devoted to conclusions.

\section{Definitions, notations and main assumption}

First of all, we set for the rest of this work the following:
\medskip

\noindent
{\bf Notation.} By using $p$, $t$, $q$, $k$ and $\Delta$, we will denote five positive integers such that $p$ is a prime number, $q=p^t$ and $\Delta=(q^k-1)/(q-1)$. From now on, $\gamma$ will denote a fixed primitive element of $\bbbf_{q^k}$. For any integer $a$, the polynomial $h_a(x) \in \bbbf_{q}[x]$ will denote the minimal polynomial of $\gamma^{-a}$. Furthermore, we will denote by ``Tr", the absolute trace mapping from $\bbbf_{q^k}$ to the prime field $\bbbf_p$, and by ``$\mbox{Tr}_{\bbbf_{q^k}/\bbbf_q}$" the trace mapping from $\bbbf_{q^k}$ to $\bbbf_q$. For any positive divisor $m$ of $q^k-1$ and for any $0 \leq i \leq m-1$, we define ${\cal D}_i^{(m)}:=\gamma^i \langle \gamma^m \rangle$, where $\langle \gamma^m \rangle$ denotes the subgroup of $\bbbf_{q^k}^*$ generated by $\gamma^m$. The cosets ${\cal D}_i^{(m)}$ are called the {\em cyclotomic classes} of order $m$ in $\bbbf_{q^k}$. In connection with these cyclotomic classes, we have the {\em cyclotomic numbers} of order $m$: 

$$(i,j)^{(m,q^k)}:=|({\cal D}_i^{(m)}+1) \cap {\cal D}_j^{(m)}| \; ,$$

\noindent
where $({\cal D}_i^{(m)}+1)=\{x+1 \:|\: x \in {\cal D}_i^{(m)} \}$, and $0 \leq i,j \leq m-1$. Finally, the canonical additive character $\chi$, of $\bbbf_{q^k}$, is:

$$\chi(y)=\zeta^{\mbox{Tr}(y)}_p \; , \;\;\;\; \mbox{ for all } y \in \bbbf_{q^k} \; ,$$

\noindent
where $\zeta_p=\exp(\frac{2\pi \sqrt{-1}}{p})$.

\medskip

The following definition is a proper generalization of the idea of a semiprimitive irreducible cyclic code that was introduced recently in \cite{diez}.

\begin{definition}\label{defuno}
Let $p$, $q$, $k$ and $\Delta$ be as before, and for any integer $a$ let $u=\gcd(\Delta,a)$. Then an irreducible cyclic code with parity-check polynomial $h_{a}(x)$, of degree $k$, is called a {\em semiprimitive code} if $u \geq 2$, and if $-1$ is a power of $p$ modulo $u$ (that is, if the prime $p$ is {\em semiprimitive modulo} $u$). 
\end{definition}

Now, we also set for the rest of this work the following:

\medskip

\noindent
{\bf Main assumption.} From now on, we are going to suppose that $3 | (q^k-1)$, $2 | \Delta$ and $3 \nmid \Delta$. Therefore, in what follows, we will reserve the Greek letter $\tau$ in order to fix $\tau=\gamma^{\frac{q^k-1}{3}}$.

\medskip

\begin{remark}\label{rmuno} 
As a consequence of our main assumption, note that $k$ should be an even integer, whereas $q$ must be an odd integer greater than $5$, and necessarily $3 | (q-1)$ and $4 | (q^k-1)$. In addition, observe that $\bbbf_{q}^* \subset {\cal D}_0^{(2)}$, $\tau \in {\cal D}_0^{(2)}$, and also that the finite field element $\tau$ is a primitive three-root of unity satisfying $\tau^2+\tau+1=0$.
\end{remark}

The following, is a well known result (\cite[Lemma 4]{seis}):

\begin{lemma}\label{lemauno}
Define

\[\eta_i = \sum_{x \in {\cal D}_i^{(2)}} \chi(x) \; , \;\;\; i=0,1 \; .\]

\noindent
Then $\eta_1=-1-\eta_0$, and

\[\eta_0 = \left\{ \begin{array}{cl}
		\frac{-1+(-1)^{tk-1}q^{k/2}}{2} & \mbox{ if } p \equiv 1 \pmod{4} \\
		\frac{-1+(-1)^{tk-1}(\sqrt{-1})^{tk}q^{k/2}}{2} & \mbox{ if } p \equiv 3 \pmod{4}
			\end{array}
\right . .\]
\end{lemma} 

The exponential sums $\eta_0$ and $\eta_1$ are known as the {\em Gaussian periods} of order $2$. Since we will be dealing with the Gaussian period of order $2$, we also need the cyclotomic numbers of order $2$. The following result is on that direction (\cite{seis}).

\begin{lemma}\label{lemados}
Suppose that $4 | (q^k-1)$, then

\[\begin{array}{rll}
	(0,0)^{(2,q^k)} & = & \frac{q^k-5}{4} \; ,\\
	(0,1)^{(2,q^k)}=(1,0)^{(2,q^k)}=(1,1)^{(2,q^k)} & = & \frac{q^k-1}{4} \; .
\end{array}\]
\end{lemma} 

\section{Some Preliminary and General Results}

With our current notation and main assumption in mind we present the following results. 

\begin{lemma}\label{lematres}
Let $a_1$ and $a_2$ be integers such that $a_1-a_2=\pm \frac{q^k-1}{3}$. Then $\gcd(\Delta,a_2)=\gcd(\Delta,a_1)$. Consequently, if $2=\gcd(\Delta,a_i)$, for $i=1,2$, then $\gcd(q^k-1,a_i)=2\frac{q-1}{\lambda_i}$, for some integer $\lambda_i$ satisfying $\gcd(q-1,\frac{a_i}{2})=\frac{q-1}{\lambda_i}$. In addition, if $3 \nmid \lambda_1$ then $\lambda_2=3\lambda_1$. Consequently, if $a$, $n$ and $\lambda$ are as in Theorem \ref{teouno} then $3 | \lambda$, $\gcd(\Delta,2 \frac{q-1}{\lambda})=2$ and $n=\lambda\frac{\Delta}{2}$.
\end{lemma} 

\begin{proof}
Clearly, $\gcd(\Delta,a_2)=\gcd(\Delta,a_1\pm\Delta\frac{(q-1)}{3})=\gcd(\Delta,a_1)$. On the other hand, for $i=1,2$, $\gcd(q^k-1,a_i)=\gcd(\Delta(q-1),a_i)=\gcd(2(q-1),a_i)=2\frac{q-1}{\lambda_i}$, for some divisor $\lambda_i$ of $q-1$ satisfying $\gcd(q-1,\frac{a_i}{2})=\frac{q-1}{\lambda_i}$.

Now, if $3 \nmid \lambda_1$ then $\frac{q-1}{\lambda_1} \nmid \frac{q-1}{3}$. Thus, since $3 \nmid \frac{\Delta}{2}$, we have $\frac{q-1}{\lambda_2}=\gcd(q-1,\frac{a_2}{2})=\gcd(q-1,\frac{a_1}{2}\pm\frac{q-1}{3}\frac{\Delta}{2})=\frac{q-1}{3\lambda_1}$. 
\end{proof}

\begin{lemma}\label{lemacuatro}
Let $\lambda$ be a divisor of $q-1$ such that $3 | \lambda$. If $\gcd(\Delta,2\frac{(q-1)}{\lambda})=2$ then, for any integer $i$,

\[\{xy \; | \; x \in {\cal D}_i^{(\frac{6(q-1)}{\lambda})} \mbox{ and } y \in \bbbf_{q}^* \}=\frac{\lambda}{3} * {\cal D}_i^{(2)} \; ,\]

\noindent
where $\frac{\lambda}{3} * {\cal D}_i^{(2)}$ is the multiset in which each element of ${\cal D}_i^{(2)}$ appears with multiplicity $\frac{\lambda}{3}$.
\end{lemma} 

\begin{proof}
Since ${\cal D}_i^{(j)}=\gamma^i {\cal D}_0^{(j)}$, for all integers $i$ and $j$, it follows that, without loss of generality, we can suppose $i=0$. 

Clearly, $\gcd(\Delta,\frac{6(q-1)}{\lambda})=2$ and $\frac{6(q-1)}{\lambda} | (q^k-1)$. Therefore, for each $x \in {\cal D}_0^{(\frac{6(q-1)}{\lambda})}$ and $y \in \bbbf_{q}^*$, there exist unique integers $l_1$ and $l_2$, with $0 \leq l_1 < \frac{\lambda\Delta}{6}$ and $0 \leq l_2 < q-1$, such that

\begin{eqnarray}
xy &=& \gamma^{\frac{6(q-1)}{\lambda}l_1+\Delta l_2}  \nonumber \\
   &=& (\gamma^{2})^{\frac{3(q-1)}{\lambda}l_1+\frac{\Delta}{2}l_2} \; . \nonumber 
\end{eqnarray}
 
\noindent
Therefore, $xy \in {\cal D}_0^{(2)}$. Now, trivially, we have $\frac{\Delta}{2} | \frac{\lambda\Delta}{6}$ and $\frac{(q-1)}{(\lambda/3)} | (q-1)$. Thus, since $\gcd(\frac{\Delta}{2},\frac{3(q-1)}{\lambda})=1$ and $|\langle \gamma^{2} \rangle|=\frac{q^k-1}{2}$, we conclude that each element $xy$ will appear with multiplicity $\frac{\lambda\Delta}{6}(q-1)/(\frac{q^k-1}{2})=\frac{\lambda}{3}$.     
\end{proof}

Now, let $\sigma$ be any fixed integer such that $3 \nmid \sigma$, and let $i,j$ be integers, with $i=0,1,2$ and $j=0,1$. By using $\tau$, and the cyclotomic classes of order $2$, we define the following sets:

\begin{eqnarray}
{\cal E}_{i,j}&=&\{(\alpha,-\tau^{i \sigma}\alpha) \in \bbbf_{q^k}^2 \;|\; (\alpha-\tau^{\sigma}\alpha) \in {\cal D}_{j}^{(2)} \}, \; \mbox{ and } \nonumber \\ 
{\cal G}&=&\{(\alpha,-\beta) \in \bbbf_{q^k}^2 \;\! | \; (\alpha-\tau^{i \sigma} \beta) \neq 0, \; 0 \leq i < 3 \:\} \; . \nonumber
\end{eqnarray}

\begin{remark}\label{rmdos}
Through a direct inspection it is easy to see that the above seven sets are pairwise disjoint and their union is equal to $\bbbf_{q^k}^2 \setminus (0,0)$. In addition, clearly $|{\cal E}_{i,j}|=|{\cal D}_{0}^{(2)}|=\frac{q^k-1}{2}$, $|{\cal G}|=q^{2k}-1-6|{\cal E}_{0,0}|=(q^k-1)(q^k-2)$ and, due to Remark \ref{rmuno}, we have that if $(\alpha-\tau^{\sigma}\alpha) \in {\cal D}_{j}^{(2)}$, then necessarily $(\alpha-\tau^{2\sigma}\alpha)=-\tau^{2\sigma} (\alpha-\tau^{\sigma}\alpha) \in {\cal D}_{j}^{(2)}$, for any integer $j=0,1$.
\end{remark} 

Now, for each $(\alpha,-\beta) \in {\cal G}$, we define the function $f_{\alpha,\beta} : \{0,1,2\} \to \{0,1\}$, given by the rule $f_{\alpha,\beta}(i)=j$ if and only if $(\alpha-\tau^{i \sigma} \beta) \in {\cal D}_{j}^{(2)}$. With the help of these functions we induce a partition of the set ${\cal G}$ into the following disjoint subsets:  

$${\cal S}_l = \{ (\alpha,-\beta) \in {\cal G} \;|\; W_H(f_{\alpha,\beta}(0),f_{\alpha,\beta}(1),f_{\alpha,\beta}(2))=l \:\}\;,$$

\noindent
for $l=0,1,2,3$, where $W_H(\cdot)$ stands for the usual Hamming weight function. 

\begin{remark}\label{rmtres}
For any $\alpha,\beta \in \bbbf_{q^k}$, we define $u_i=(\alpha-\tau^{i \sigma} \beta)$, for $i=0,1,2$. It is not difficult to see that these $u$ values satisfy $u_0 + u_1\tau^{\sigma} + u_2\tau^{2\sigma} = 0$. Furthermore, observe that if we arbitrarily choose the values of, say, $u_0$ and $u_2$ then there must exist a unique vector $(\alpha,\beta) \in \bbbf_{q^k}^2$, such that $u_0=(\alpha-\beta)$, $u_2=(\alpha-\tau^{2\sigma} \beta)$ and $u_1=-\tau^{-\sigma}(u_0+u_2\tau^{2\sigma})$. Therefore, if we want to calculate, for example, $|{\cal S}_0|$ then we can assume, without loss of generality, that $u_2$ can take any value in ${\cal D}_{0}^{(2)}$. This leads us to $\frac{q^k-1}{2}$ possible choices for $u_2$. But $u_1=-u_2\tau^{\sigma}(\frac{u_0}{u_2}\tau^{-2\sigma}+1)$ and $-1, \tau \in {\cal D}_{0}^{(2)}$ (see Remark \ref{rmuno}), thus, in order that $u_1$ and $u_0$ also belong to ${\cal D}_{0}^{(2)}$ it is necessary that $(\frac{u_0}{u_2}\tau^{-2\sigma}+1) \in {\cal D}_{0}^{(2)}$. Hence, the number of such instances is given by the cyclotomic number $(0,0)^{(2,q^k)}$. Consequently, we have $|{\cal S}_0|=\frac{q^k-1}{2}(0,0)^{(2,q^k)}$.  In a quite similar way, one can obtain $|{\cal S}_1|$, $|{\cal S}_2|$ and $|{\cal S}_3|$.   
\end{remark} 

Keeping in mind the previous definitions and observations we now present the following result, which will be important in order to determine the weight distribution of the class of reducible cyclic codes that we are interested in.

\begin{lemma}\label{lemacinco}
With our notation and main assumption, we have 

\begin{eqnarray}
|{\cal S}_0|&=&\frac{q^k-1}{2}(0,0)^{(2,q^k)} \nonumber \\
|{\cal S}_1|&=&\frac{3(q^k-1)}{2}(0,1)^{(2,q^k)} \nonumber \\
|{\cal S}_2|&=&\frac{3(q^k-1)}{2}(1,1)^{(2,q^k)} \nonumber \\
|{\cal S}_3|&=&(q^k-1)(q^k-2)-(|{\cal S}_0|+|{\cal S}_1|+|{\cal S}_2|) \; . \nonumber
\end{eqnarray}

\noindent
Furthermore, if $\chi$ denotes the canonical additive character of $\bbbf_{q^k}$, and if $\eta_0$ and $\eta_1$ are as in Lemma \ref{lemauno}, and by considering $\sigma$ as before, then, for any $\alpha,\beta \in \bbbf_{q^k}$, we also have

\[\sum_{z \in {\cal D}_{0}^{(2)}} \sum_{i=0}^{2} \chi(z(\alpha+\tau^{i \sigma}\beta)) \!=\! \left\{ \begin{array}{cl}
		\!\! \frac{3(q^k-1)}{2} & \!\! \mbox{ if } (\alpha,\beta)\!=\!(0,0) \\
		\!\! \frac{q^k-1}{2}+2\eta_0 & \!\! \mbox{ if } (\alpha,\beta) \in {\cal W}_0 \\
		\!\! \frac{q^k-1}{2}+2\eta_1 & \!\! \mbox{ if } (\alpha,\beta) \in {\cal W}_1 \\
		\!\! 3\eta_0 & \!\! \mbox{ if } (\alpha,\beta) \in {\cal S}_0 \\
		\!\! -1+\eta_0 & \!\! \mbox{ if } (\alpha,\beta) \in {\cal S}_1 \\
		\!\! -1+\eta_1 & \!\! \mbox{ if } (\alpha,\beta) \in {\cal S}_2 \\
		\!\! 3\eta_1 & \!\! \mbox{ if } (\alpha,\beta) \in {\cal S}_3 \: , 
			\end{array}
\right . \]

\noindent
where ${\cal W}_0=\cup_{i=0}^2 {\cal E}_{i,0}$ and  ${\cal W}_1=\cup_{i=0}^2 {\cal E}_{i,1}$.
\end{lemma} 

\begin{proof}
The first assertion comes from Remark \ref{rmtres}. Since $\sum_{z \in {\cal D}_0^{(2)}} \chi(0)=|{\cal D}_0^{(2)}|=\frac{q^k-1}{2}$, the second assertion comes from Lemma \ref{lemauno}, Remark \ref{rmdos}, and from the definitions of the sets ${\cal E}_{i,j}$ and ${\cal S}_l$, with $i=0,1,2$, $j=0,1$ and $l=0,1,2,3$.  
\end{proof}

Considering the actual values of the cyclotomic numbers in Lemma \ref{lemados}, the following result is an important consequence.

\begin{center}
TABLE III \\
{\em Value distribution of ${\displaystyle \sum_{z \in {\cal D}_{0}^{(2)}} \sum_{i=0}^{2} \chi(z(\alpha+\tau^{i \sigma}\beta))}$.}
\end{center}
\begin{center}
\begin{tabular}{|c|c|} \hline
{\bf Value} & $\;$ {\bf Frequency} $\;$\\ \hline \hline
$\frac{3(q^k-1)}{2}$ & 1 \\ \hline
$\;\;\; \frac{q^k-1}{2}+2\eta_0 \;\;\;$ & $\frac{3(q^k-1)}{2}$ \\ \hline
$\;\;\; \frac{q^k-1}{2}+2\eta_1 \;\;\;$ & $\frac{3(q^k-1)}{2}$ \\ \hline
$\; 3\eta_0 \;$ & $\frac{(q^k-1)(q^k-5)}{8}$ \\ \hline
$-1+\eta_0$ & $\frac{3(q^k-1)^2}{8}$ \\ \hline
$-1+\eta_1$ & $\frac{3(q^k-1)^2}{8}$ \\ \hline
$\; 3\eta_1 \;$ & $\frac{(q^k-1)(q^k-5)}{8}$ \\ \hline 
\end{tabular}
\end{center}

\begin{corollary}\label{coruno}
Consider the same hypotheses as in the previous lemma. Then the value distribution of the character sum $\sum_{z \in {\cal D}_{0}^{(2)}} \sum_{i=0}^{2} \chi(z(\alpha+\tau^{i \sigma}\beta))$ is given in Table III. 
\end{corollary}

\section{Formal proof of Theorem 1}

\begin{proof}
Part {\em (A)}: Let $\lambda_1$ and $\lambda_2$ be as in Lemma \ref{lematres}, and, for $i=1,2$, let $v_i$ be a positive integer such that $a_i q^{v_i} \equiv a_i \pmod{q^k-1}$. Then $(q^k-1) | a_i(q^{v_i} -1)$, but due to Lemma \ref{lematres} we know that $\gcd(q^k-1,a_i)=2 \frac{q-1}{\lambda_i}$. In consequence, we necessarily have $(q^k-1) | 2 \frac{q-1}{\lambda_i} (q^{v_i} -1)$, which in turn implies that $\frac{\Delta}{2} | (\frac{q-1}{\lambda_i}) (\frac{q^{v_i} -1}{q-1})$. Now, $\gcd(\frac{\Delta}{2},\frac{q-1}{\lambda_i})=1$, thus $(q^k -1)|2(q^{v_i}-1)$. However, this last condition is impossible if ${v_i}<k$ and $q \geq 7$ (recall Remark \ref{rmuno}). Hence $\deg(h_{a_i}(x))=k$.

Now, suppose that $h_{a_1}(x) = h_{a_2}(x)$. Then, there exists an integer $0 \leq v < k$ such that $a_1 q^v \equiv a_2 \! \pmod{q^k-1}$. But $a_2 = a_1 \pm \frac{q^k-1}{3}$, thus the last congruence implies $a_1(q^v-1) \equiv \pm \frac{q^k-1}{3} \pmod{q^k-1}$, which in turn implies that $a_1(q^v-1) \equiv 0 \pmod{\frac{q^k-1}{3}}$. That is, $(q^k-1) | 3a_1(q^v-1)$. But $\gcd(q^k-1,a_1)=2 \frac{q-1}{\lambda_1}$, thus $(q^k-1) | 6 (\frac{q-1}{\lambda_1}) (q^v -1)$, consequently $\frac{\Delta}{2} | 3(\frac{q-1}{\lambda_1}) (\frac{q^v -1}{q-1})$. Now, $\gcd(\frac{\Delta}{2},\frac{q-1}{\lambda_1})=1$ therefore $(q^k -1)|6(q^v-1)$. However, this last condition is impossible if $v<k$ and $q \geq 7$ (recall Remark \ref{rmuno}). Hence $h_{a_1}(x) \neq h_{a_2}(x)$.

Due to Lemma \ref{lematres}, we know that $n=\lambda\frac{\Delta}{2}$, with $3 | \lambda$. But $\gcd(q^k-1,a)=2 \frac{q-1}{\lambda}$, thus $an \equiv 0 \pmod{q^k-1}$ and $(a \pm \frac{q^k-1}{3})n=an \pm (q^k-1)\frac{n}{3} \equiv 0 \pmod{q^k-1}$. Therefore the three cyclic codes ${\cal C}_{(a_1)}$, ${\cal C}_{(a_2)}$ and ${\cal C}_{(a_1,a_2)}$ have length $n$.

Since $\gcd(\Delta,a_1)=\gcd(\Delta,a_2)=2$ (Lemma \ref{lematres} again), clearly, in accordance with Definition \ref{defuno}, ${\cal C}_{(a_1)}$ and ${\cal C}_{(a_2)}$ are two different semiprimitive two-weight irreducible cyclic codes. Thus, by means of the characterization for this kind of codes in \cite[Theorem 7]{diez}, we can see that their weight distributions are as is shown in Table I.

Part {\em (B)}: First of all, observe that due to Lemma \ref{lematres}, $3 | \lambda$. Furthermore, since ${\cal C}_{(a_2,a_1)}={\cal C}_{(a_1,a_2)}$, we can assume without loss of generality that $a_1=a$ and $a_2=a \pm \frac{q^k-1}{3}$.

Clearly, the cyclic code ${\cal C}_{(a_1,a_2)}$ has length $n$ and its dimension is $2k$ due to Part {\em (A)}.

Now, for each $\alpha,\beta \in \bbbf_{q^k}$, we define $c(n,a_1,a_2,\alpha,\beta)$ as the vector of length $n$ over $\bbbf_q$, which is given by:

$$(\mbox{Tr}_{\bbbf_{q^k}/\bbbf_q}(\alpha(\gamma^{a_1})^i+\beta(\gamma^{a_2})^i))_{i=0}^{n-1}\; .$$


\noindent
Thanks to Delsarte's Theorem (\cite{uno}), it is well known that 

$${\cal C}_{(a_1,a_2)}=\{ c(n,a_1,a_2,\alpha,\beta) \: | \: \alpha,\beta \in \bbbf_{q^k} \} \; .$$ 

\noindent
Thus the Hamming weight of any codeword $c(n,a_1,a_2,\alpha,\beta)$ is equal to $n-Z(\alpha,\beta)$, where

$$Z(\alpha,\beta)\!=\!\sharp\{\;i\; | \; \mbox{Tr}_{\bbbf_{q^k}/\bbbf_q}(\alpha\gamma^{a_1 i}+\beta\gamma^{a_2 i})=0, \: 0 \leq i < n \} \:.$$


Now, if $\chi'$ is the canonical additive character of $\bbbf_q$, then, by the orthogonal property of $\chi'$ (see, for example, \cite[p. 192]{cinco}), we know that for each $c \in \bbbf_q$ we have

\[\sum_{y \in \bbbf_q} \chi'(yc)=
\left\{ \begin{array}{ll}
		q & \mbox{ if $c=0$} \\
		0 & \mbox{ if $c\neq0$}
			\end{array}
\right . \; ,\]

\noindent
thus

\[Z(\alpha,\beta)=\frac{1}{q}\sum_{i=0}^{n-1} \sum_{y \in \bbbf_q} \chi'(\mbox{Tr}_{\bbbf_{q^k}/\bbbf_q}(y(\alpha\gamma^{a_1 i}+\beta\gamma^{a_2 i}))) \;. \]

\noindent
If $\chi$ denotes the canonical additive character of $\bbbf_{q^k}$, then $\chi'(\mbox{Tr}_{{\bbbf_{q^k}/\bbbf_q}}(\varepsilon))=\chi(\varepsilon)$ for all $\varepsilon \in \bbbf_{q^k}$. Therefore, we have

\[Z(\alpha,\beta)=\frac{n}{q}+\frac{1}{q}\sum_{i=0}^{n-1} \sum_{y \in \bbbf_q^*} \chi(y(\alpha\gamma^{a_1 i}+\beta\gamma^{a_2 i})) \; .\]

Now, since $\gcd(q^k-1,a_1)=2\frac{q-1}{\lambda}$, there must be an integer $v$ such that $3 \nmid v$ and $\gamma^{v a_1}=\gamma^{2\frac{q-1}{\lambda}}$. Thus, since $\tau=\gamma^{\frac{q^k-1}{3}}$ and by taking $\epsilon$ to be the integer whose value is defined by

\[\epsilon=
\left\{ \begin{array}{ll}
		1 & \mbox{ if $+$ is the sign chosen in $a_2=a_1 \pm \frac{q^k-1}{3}$} \\
		2 & \mbox{ otherwise}
			\end{array}
\right . \; ,\] 

\noindent
we have

\begin{eqnarray} 
Z(\alpha,\beta)&=&\frac{n}{q}+\frac{1}{q} \sum_{i=0}^{n-1} \sum_{y \in \bbbf_q^*} \chi(\gamma^{a_1 i}y(\alpha+\tau^{i\epsilon}\beta)) \nonumber \\ 
&=&\frac{n}{q}+\frac{1}{q}\sum_{i=0}^{n-1} \sum_{y \in \bbbf_q^*} \chi(\gamma^{2 \frac{q-1}{\lambda} i}y(\alpha+\tau^{iv \epsilon}\beta)) \; .\nonumber  
\end{eqnarray}

\noindent
But clearly $3 | n$, thus

\begin{eqnarray}
\{\gamma^{2\frac{q-1}{\lambda} i} \; | \; 0 \leq i < n \}&\!\!\!\!\!=\!\!\!\!\!&{\cal D}_{0}^{(\frac{2(q-1)}{\lambda})}   \nonumber \\
&\!\!\!\!\!=\!\!\!\!\!&{\cal D}_{0}^{(\frac{6(q-1)}{\lambda})} \cup {\cal D}_{\frac{2(q-1)}{\lambda}}^{(\frac{6(q-1)}{\lambda})} \cup {\cal D}_{\frac{4(q-1)}{\lambda}}^{(\frac{6(q-1)}{\lambda})} .\nonumber
\end{eqnarray}    

\noindent
Therefore,

\[Z(\alpha,\beta)=\frac{n}{q}+\frac{1}{q} \sum_{i=0}^{2} \sum_{x \in {\cal D}_{\frac{2(q-1)}{\lambda}i}^{(\frac{6(q-1)}{\lambda})}} \sum_{y \in \bbbf_q^*} \chi(xy(\alpha+\tau^{iv \epsilon}\beta))  \; .\] 

\noindent
Now, we already said that $3 | \lambda$ and $\gcd(\Delta,2\frac{q-1}{\lambda})=2$, thus after applying Lemma \ref{lemacuatro}, we obtain 

\[Z(\alpha,\beta)=\frac{n}{q}+\frac{\lambda}{3q} \sum_{i=0}^{2} \sum_{z \in {\cal D}_{\frac{2(q-1)}{\lambda}i}^{(2)}} \chi(z(\alpha+\tau^{iv \epsilon}\beta)) \; .\]

\noindent
But, $2 | \frac{2(q-1)}{\lambda}$, thus

\[Z(\alpha,\beta)=\frac{n}{q}+\frac{\lambda}{3q} \sum_{z \in {\cal D}_{0}^{(2)}} \sum_{i=0}^{2} \chi(z(\alpha+\tau^{iv \epsilon}\beta)) \; .\]

\noindent
Clearly $3 \nmid (v \epsilon)$, therefore the result comes from Corollary \ref{coruno} because the Hamming weight of any codeword of the form $c(n,a_1,a_2,\alpha,\beta)$ in ${\cal C}_{(a_1,a_2)}$ is equal to $n-Z(\alpha,\beta)$.  
\end{proof}

As was mentioned before, Theorem \ref{teouno} deals with the kind of reducible cyclic codes whose parity check polynomials are given by the products of the form $h_{a}(x)h_{a \pm \frac{q^k-1}{3}}(x)$, where $a$ is any integer and $h_{a}(x) \neq h_{a \pm \frac{q^k-1}{3}}(x)$. The following result, which is the main result in \cite[Theorem 3.6]{once}, also deals with this kind of reducible cyclic codes:

\begin{center}
TABLE IV \\
{\em Weight distribution of ${\cal C}_{(\frac{q-1}{h},\frac{q-1}{h}+\frac{q^k-1}{3})}$.}
\end{center}
\begin{center}
\begin{tabular}{|c|c|} \hline
{\bf Weight} & $\;$ {\bf Frequency} $\;$\\ \hline \hline
0 & 1 \\ \hline
$\; \frac{2h}{3} (q^{k-1}-q^{(k-2)/2}) \;$ & $\frac{3(q^k-1)}{2}$ \\ \hline
$\; \frac{2h}{3} (q^{k-1}+q^{(k-2)/2}) \;$ & $\frac{3(q^k-1)}{2}$ \\ \hline
$\; h (q^{k-1}-q^{(k-2)/2}) \;$ & $\frac{(q^k-1)(q^k-5)}{8}$ \\ \hline
$\; h (q^{k-1}+q^{(k-2)/2}) \;$ & $\frac{(q^k-1)(q^k-5)}{8}$ \\ \hline
$\; \frac{h}{3} (3q^{k-1}-q^{(k-2)/2}) \;$ & $\frac{3(q^k-1)^2}{8}$ \\ \hline
$\; \frac{h}{3} (3q^{k-1}+q^{(k-2)/2}) \;$ & $\frac{3(q^k-1)^2}{8}$ \\ \hline 
\end{tabular}
\end{center}

\begin{theorem}\label{teodos}
Let $h$ be a positive factor of $q-1$ such that $3 | h$. If $\gcd(k,3(q-1)/h)=2$, then ${\cal C}_{(\frac{q-1}{h},\frac{q-1}{h}+\frac{q^k-1}{3})}$ is an $[h\Delta,2k]$ code with the weight distribution in Table IV.
\end{theorem}

The following result shows that the family of codes in Theorem \ref{teodos}, is included in Theorem \ref{teouno}.

\begin{theorem}\label{teotres}
Conditions in Theorem \ref{teodos} imply conditions in Theorem \ref{teouno}.
\end{theorem}

\begin{proof}
Clearly $3 | (q^k-1)$, and because $\gcd(\Delta,\rho)=\gcd(k,\rho)$ for all $\rho | (q-1)$ (see, for example \cite[Remark 3]{ocho}), we have $\gcd(\Delta,3(q-1)/h)=2$, $2 | \Delta$, and $3 \nmid \Delta$. Thus, by taking $a=(q-1)/h$ in Theorem \ref{teouno}, the result follows directly. 
\end{proof}

The following are direct applications of Theorem \ref{teouno}.

\begin{example}\label{ejecero}
Let $q=13$, $k=2$, $a_1=8$ and $a_2=a_1+\frac{q^k-1}{3}=64$. Then, $3 | (q^k-1)$, $\Delta=14$, $3 \nmid \Delta$, $n=21$, $a=8$ and $\lambda=3$. By Theorem \ref{teouno}, ${\cal C}_{(8)}$ and ${\cal C}_{(64)}$ are two different semiprimitive two-weight irreducible cyclic codes of length $21$, dimension $2$ and weight enumerator polynomial $A(z)=1+84z^{18}+84z^{21}$. In addition, ${\cal C}_{(8,64)}$ is a cyclic code of length $21$, dimension $4$ and weight enumerator polynomial 

\begin{eqnarray}
A(z)&=&1+252z^{12}+252z^{14}+3444z^{18}+ \nonumber \\
&&10584z^{19}+10584z^{20}+3444z^{21} \; . \nonumber
\end{eqnarray}
\end{example}

\begin{remark}
Since $2 | \Delta$, clearly the length of all codes in Theorem \ref{teodos} must be an even number. Therefore, the code in the previous example does not belong to the class of codes in Theorem \ref{teodos}.
\end{remark}

\begin{example}\label{ejeuno}
Let $q=7$, $k=2$, $a_1=2$ and $a_2=a_1-\frac{q^k-1}{3}=-14$. Then, $3 | (q^k-1)$, $\Delta=8$, $3 \nmid \Delta$, $n=24$, $a=2$ and $\lambda=6$. By Theorem \ref{teouno}, ${\cal C}_{(2)}$ and ${\cal C}_{(-14)}$ are two different semiprimitive two-weight irreducible cyclic codes of length $24$, dimension $2$ and weight enumerator polynomial $A(z)=1+24z^{18}+24z^{24}$. In addition, ${\cal C}_{(2,-14)}$ is a cyclic code of length $24$, dimension $4$ and weight enumerator polynomial 

\begin{eqnarray}\label{equltima}
A(z)&=&1+72z^{12}+72z^{16}+264z^{18}+ \nonumber \\
&&864z^{20}+864z^{22}+264z^{24} \; .
\end{eqnarray}
\end{example}

\begin{remark}
Suppose again that $q=7$ and $k=2$. Then $h_{-14}(x)=h_{34}(x) \neq h_{18}(x)$, and consequently note that despite that the weight enumerator polynomial in the previous example is exactly the same as weight enumerator polynomial in the example of \cite[page 7257]{dos}, the cyclic codes ${\cal C}_{(2,-14)}$ and ${\cal C}_{(2,18)}$ are different. In addition, also note that the cyclic code ${\cal C}_{(2,-14)}={\cal C}_{(2,34)}$ does not belong to the class of codes in Theorem \ref{teodos}.
\end{remark}

\begin{remark}
Through a direct inspection, it is not difficult to see that all different reducible cyclic codes over $\bbbf_{7}$ of length $24$, dimension $4$ and weight enumerator polynomial, as in (\ref{equltima}), are ${\cal C}_{(2,18)}$,  ${\cal C}_{(2,34)}$, ${\cal C}_{(18,34)}$, ${\cal C}_{(6,10)}$, ${\cal C}_{(6,26)}$ and ${\cal C}_{(10,26)}$. All these reducible cyclic codes belong to the class of codes in Theorem \ref{teouno}. 
\end{remark}

\begin{remark}
Under our main assumption and due to Lemma \ref{lematres}, note that a reducible cyclic code, ${\cal C}_{(a,a \pm \frac{q^k-1}{3})}$, will belong to the family of codes in Theorem \ref{teouno}, if and only if $\gcd(\Delta,a)=2$. This condition, which is easy to verify, allows us to identify all the reducible cyclic codes that satisfy conditions in Theorem \ref{teouno}. In fact, if ${\cal N}_{(q,k)}$ is the number of different reducible cyclic codes, ${\cal C}_{(a_1,a_2)}$, that satisfy such conditions, then it is not difficult to see that

$${\cal N}_{(q,k)}=\frac{\phi(\frac{\Delta}{2})(q-1)}{k}\;,$$ 

\noindent
where $\phi$ is the usual Euler $\phi$-function. What is interesting here is that it seems that ${\cal N}_{(q,k)}$ is also the total number of reducible cyclic codes whose weight distribution is given in Table II.
\end{remark}

\section{Conclusion}
In this work we found the sufficient numerical conditions in order to obtain the full weight distribution of a family of codes that belongs to the kind of reducible cyclic codes whose parity check polynomials are given by the products of the form $h_{a}(x)h_{a \pm \frac{q^k-1}{3}}(x)$. By means of the characterization of all semiprimitive two-weight irreducible cyclic codes that was presented in \cite[Theorem 7]{diez}, we were able to identify that the codes in this family are constructed as the direct sum of two different semiprimitive two-weight irreducible cyclic codes. In addition, we also showed that the class of codes recently studied in \cite{once} is included in this family. Moreover, despite of what was stated in \cite{once}, we showed that, at least for the codes in this family, it is still possible to compute the frequencies of their weight distributions through the cyclotomic numbers in a very easy way. 

Finally, we believe that perhaps, following the same idea as in [9], it could be possible to develop a more general theory that allows us to present a unified explanation for an enlarged family of reducible cyclic codes of this kind.

\bibliographystyle{IEEE}

\medskip
\medskip


\end{document}